\documentclass[reprint,superscriptaddress,amsmath, amssymb, amsfonts,preprintnumbers,aps,prl,longbibliography,
nofootinbib, nobibnotes
]{revtex4-1}

\allowdisplaybreaks

\usepackage{graphicx}% Include figure files
\usepackage{dcolumn}% Align table columns on decimal point
\usepackage{bm,amstext,amssymb,amsmath}% math

\usepackage{graphicx,color}
\usepackage{delimset} % for brackets
\usepackage[colorlinks=true]{hyperref}
\usepackage{orcidlink}
\usepackage[dvipsnames]{xcolor}
\usepackage{tikz}
\usepackage{verbatim}
\usetikzlibrary{decorations.pathmorphing}
\usetikzlibrary{decorations.markings}
\usetikzlibrary{patterns}
\tikzset{
	compobj/.style={draw=black},
	binary/.style={draw=black, double distance=0.03cm},
	potgrav/.style={draw=black},
	phigrav/.style={draw=blue},
	Agrav/.style={draw=red},
	sigmagrav/.style={draw=LimeGreen},
	radgrav/.style={decorate, draw=black, segment length=5pt, decoration={snake,amplitude=2pt} },
	radgravlarge/.style={decorate,segment length=20pt, decoration={snake,amplitude=2pt}, draw},
	partial ellipse/.style args={#1:#2:#3}{
        insert path={+ (#1:#3) arc (#1:#2:#3)}},
    blackdot/.style={
		circle,
		draw=black,
		fill=black,
		inner sep=0pt,
		minimum size=0.3ex
	}
}
\tikzstyle directed=[postaction={decorate,decoration={markings,
		mark=at position .25 with {\arrow{stealth}},
		mark=at position .75 with {\arrow{stealth}}}}]

\newcommand{\dd}{\mathrm{d}}

\newcommand{\cV}{\mathcal{V}}
\newcommand{\ioverhbar}{\frac{\ii}{\hbar}}
\newcommand{\pnord}{k}

\def\ii{ {\bf i}}

\newcommand{\PRLsection}[1]{\textit{#1} ---}
\newcommand{\PRLsubsection}[1]{\textit{#1}: }

\DeclareFontFamily{OT1}{pzc}{}
\DeclareFontShape{OT1}{pzc}{m}{it}{<-> s * [1.350] pzcmi7t}{}
\DeclareMathAlphabet{\mathpzc}{OT1}{pzc}{m}{it}

\definecolor{darkred}{rgb}{0.5,0.0,0.0}
\hypersetup{
  colorlinks=true,
  linkcolor=darkred,   % internal links (sections, equations)
  citecolor=darkred,   % bibliography citations
  urlcolor=darkred     % URLs
}

\begin{document}

\preprint{HU-EP-26/15-RTG}

\title{
All-order structure of static gravitational interactions \\ 
and the seventh post-Newtonian potential
}

\newcommand{\aei}{Max Planck Institute for Gravitational Physics (Albert Einstein Institute), D-14476 Potsdam, Germany}
\newcommand{\hu}{Institut f{\"u}r Physik und IRIS Adlershof, Humboldt-Universit {\"a}t zu Berlin, Zum Großen Windkanal 2, D-12489 Berlin, Germany}
\newcommand{\sns}{Scuola Normale Superiore, Piazza dei Cavalieri 7, 56126, Pisa, Italy}
\newcommand{\infnpi}{INFN Sezione di Pisa, Largo
Pontecorvo 3, 56127 Pisa, Italy}
\newcommand{\edi}{Higgs Centre for Theoretical Physics, University of Edinburgh, James Clerk Maxwell Building,Peter Guthrie Tait Road, Edinburgh, EH9 3FD, United Kingdom }
\newcommand{\unipd}{Dipartimento di Fisica e Astronomia, Universita di Padova, Via Marzolo 8, 35131 Padova, Italy}
\newcommand{\infnpd}{INFN, Sezione di Padova,
Via Marzolo 8, I-35131 Padova, Italy.}

\author{Giacomo Brunello\,\orcidlink{0009-0004-4788-738X}}
\email{giacomo.brunello@sns.it}
\affiliation{\sns}
\affiliation{\infnpi}

\author{Manoj K. Mandal\, 
\orcidlink{0000-0003-0850-7685}}
\email{manojkumar.mandal@unipd.it}
\affiliation{\unipd}
\affiliation{\infnpd}

\author{Pierpaolo Mastrolia\,\orcidlink{0000-0001-9711-7798}}
\email{pierpaolo.mastrolia@unipd.it}
\affiliation{\unipd}
\affiliation{\infnpd}

\author{Raj Patil\,\orcidlink{0000-0002-7055-0345}}
\email{raj.patil@aei.mpg.de}
\affiliation{\aei}
\affiliation{\hu}

\author{\\Matteo Pegorin\,~\orcidlink{0009-0003-1248-871X}}
\email{matteo.pegorin.1@phd.unipd.it}
\affiliation{\unipd}
\affiliation{\infnpd}
\affiliation{\aei}

\author{Sid Smith\,\orcidlink{0009-0007-7799-0136}}
\email{sid.smith@ed.ac.uk}
\affiliation{\unipd}
\affiliation{\infnpd}
\affiliation{\edi}

\author{Jan Steinhoff\,\orcidlink{0000-0002-1614-0214}}
\email{jan.steinhoff@aei.mpg.de}
\affiliation{\aei}

\begin{abstract}
We present a closed formula for the computation of static post-Newtonian corrections 
to the two-body gravitational dynamics at any odd order, assuming the lower-order results are known. The formula is derived within a correlation function framework and exploits 
the $\mathbb{Z}_2$ symmetry of the static sector, leading to a novel theoretical interpretation 
of the factorization theorem. As an application, we compute the gravitational interaction of two 
compact coalescing objects at the seventh post-Newtonian order in the static limit, which receives contributions from seven-loop graphs at order
 $\mathcal{O}(G_N^8 v^0)$, and find complete agreement with the results obtained using the diagrammatic approach of the factorization theorem.
\end{abstract}

\maketitle

\PRLsection{Introduction}
%\section{Introduction} 
The gravitational dynamics of coalescing binary systems is one of the most compelling problems in 
fundamental physics. It underlies the generation of gravitational waves (GWs), which encode information 
about the internal structure of compact objects before and after coalescence. 
Precise theoretical predictions for such systems offer a unique opportunity to 
understand the distributions of black hole masses and spins~\cite{LIGOScientific:2025pvj}, constraints on the neutron-star equation-of-state~\cite{LIGOScientific:2018cki}, measurements of the Hubble-Lema\^itre parameter~\cite{LIGOScientific:2017adf,LIGOScientific:2021aug,LIGOScientific:2025jau}, and constraints on the theory of General Relativity (GR)~\cite{LIGOScientific:2016lio,LIGOScientific:2020tif,LIGOScientific:2021sio,LIGOScientific:2025rid}. Systematic biases are already visible in the current gravitational-wave analyses~\cite{Owen:2023mid,Dhani:2024jja,LIGOScientific:2025rsn}, which call for ever more precise theoretical templates of the binary dynamics and emitted waveform.

The modeling of compact binary dynamics relies on
a combination of complementary approaches.
Numerical relativity~\cite{Pretorius:2005gq,Campanelli:2005dd,Baker:2005vv} 
is required in the strongly nonlinear merger regime, whereas the inspiral phase admits a systematic 
analytic treatment within the post-Newtonian (PN) expansion~\cite{lorentzanddroste,Einstein:1938yz,
Futamase:2007zz,Blanchet:2013haa,Porto:2016pyg,Schafer:2018kuf,Levi:2018nxp,Jaranowski:1997ky,
Damour:2014jta,Jaranowski:2015lha,Bernard:2015njp,Bernard:2016wrg,Damour:2016abl,
Blanchet:2023sbv,Blanchet:2023bwj}, 
valid for slowly moving and widely separated bodies. For bound binaries, observables are organized as 
a series in powers of $v/c$ and as a result, at $\pnord$PN order one encounters contributions of the form
$G_N^l (v^2)^{\pnord+1-l}, $ with $ 0 \le l \le \pnord+1 ,
$ where $G_N$ is the Newton constant.
The post-Minkowskian approximation~\cite{Westpfahl:1979gu,Westpfahl:1980mk,Bel:1981be,Westpfahl:1985tsl,schafer1986adm,Ledvinka:2008tk,Buonanno:2022pgc,Travaglini:2022uwo,Bjerrum-Bohr:2022blt,Kosower:2022yvp,Damour:2016gwp,
Cheung:2018wkq,Bern:2019nnu,Bern:2022kto,Bern:2024adl,Kosower:2018adc,Bjerrum-Bohr:2018xdl,Cristofoli:2019neg,Damgaard:2019lfh,Brandhuber:2021eyq,Vines:2017hyw,Damgaard:2024fqj,Mougiakakos:2024nku,Mougiakakos:2024lif,Kalin:2020mvi,Kalin:2020fhe,Mogull:2020sak,Jakobsen:2021zvh,Jakobsen:2022psy,Driesse:2024xad,Kalin:2022hph,Driesse:2024feo,Bern:2025zno,Dlapa:2025biy,Kalin:2019rwq,Dlapa:2024cje, Bern:2025wyd, Driesse:2026qiz,
Akpinar:2025bkt,Bjerrum-Bohr:2026fhx} and the gravitational self-force (SF) framework~\cite{Mino:1996nk,Quinn:1996am,Barack:2001gx,Barack:2002mh,Gralla:2008fg,Detweiler:2008ft,Keidl:2010pm,vandeMeent:2017bcc,Pound:2012nt,Pound:2019lzj,Gralla:2021qaf,Pound:2021qin,Warburton:2021kwk,Wardell:2021fyy} 
further extend the analytic description to weak-field configurations and 
small mass ratios. The most commonly used approaches to build complete inspiral-merger-ringdown  waveform models of compact binaries are the phenomenological methods~\cite{Ajith:2007qp,Pratten:2020ceb,Estelles:2021gvs} and effective-one-body formalism~\cite{Buonanno:1998gg,Buonanno:2000ef,Damour:2000we,Damour:2001tu,Buonanno:2005xu}.

The PN program began with the 1PN results of Refs.~\cite{lorentzanddroste,Einstein:1938yz}, 
and higher-order calculations have required increasingly sophisticated analytical and 
computational techniques. 
The current state of the art includes conservative dynamics at 
4PN order~\cite{Damour:2014jta,Jaranowski:2015lha,Bernard:2015njp,Bernard:2016wrg,Damour:2016abl}
and radiation at 4.5PN order~\cite{Blanchet:2023sbv,Blanchet:2023bwj}. 
Pushing beyond this frontier is necessary to match the accuracy demanded by current and future 
gravitational-wave detectors~\cite{Purrer:2019jcp,Hu:2022rjq,Owen:2023mid,LIGOScientific:2025rsn}.

Major progress in the past decade has been achieved within the effective field theory 
(EFT) approach~\cite{Goldberger:2004jt},
which reformulates the PN expansion in terms of classical Feynman 
diagrams for conservative dynamics~\cite{Foffa:2012rn,Kol:2009mj,Foffa:2019rdf,Foffa:2016rgu,Foffa:2019yfl,Blumlein:2020pog,Foffa:2019hrb,
Blumlein:2021txe,Porto:2024cwd,Blumlein:2021txj,Blumlein:2020znm,Blumlein:2019zku,
Foffa:2020nqe,Blumlein:2020pyo,
Porto:2005ac,Levi:2015msa,Kim:2021rfj,Levi:2020uwu,Levi:2019kgk,Kim:2022pou,Kim:2022bwv,
Levi:2022dqm,Levi:2022rrq,Mandal:2022nty,Mandal:2022ufb}
and radiative emission~\cite{Goldberger:2009qd,
Ross:2012fc,
Cho:2021mqw,Cho:2022syn,Amalberti:2023ohj,Amalberti:2024jaa,Mandal:2024iug}. 
Combined with modern multi-loop methods~\cite{Kol:2013ega,Foffa:2016rgu}, 
this framework led to the complete 
conservative dynamics at 4PN~\cite{Foffa:2016rgu, Foffa:2019rdf, Foffa:2019yfl}, 
and recently at 5PN~\cite{Foffa:2019hrb,Blumlein:2019zku,Foffa:2020nqe,Blumlein:2020pyo,Blumlein:2021txe,Almeida:2021xwn,Porto:2024cwd,Porto:2026fsd}. At 6PN, the order $O(G_N^4)$ contributions~\cite{Blumlein:2021txj}, partial radiative contributions~\cite{Almeida:2026clf}, and further local and nonlocal dynamical information~\cite{Bini:2020nsb, Bini:2020rzn, Bini:2020uiq, Bini:2020hmy} have been computed.

In this approach, PN calculations are equivalent 
to evaluating multi-loop massless two-point Feynman integrals, with the static sector 
corresponding to the highest-loop diagrams at each PN order.
Very recently, this perspective enabled us to compute the static two-body effective interaction potential at the
order ${\cal O}(G_N^7 v^0)$(6PN)~\cite{Brunello:2025gpf}, using the development of novel techniques and computational tools for high-loop integrals.

In this Letter, we report two main results. First, we derive a closed formula for static PN corrections at arbitrary PN order in terms of gravitational correlation functions. This leads to a novel theoretical interpretation for the odd-PN diagrammatic factorization theorem observed in Ref.~\cite{Foffa:2019hrb} as a consequence of an accidental symmetry of the static sector. Second, we obtain the first result for the static two-body effective interaction at order ${\cal O}(G_N^8 v^0)$, corresponding to 7PN. The 7PN result is derived independently using both the earlier diagrammatic factorization and the present formalism, and the two computations agree exactly. In addition to extending the PN-EFT frontier to seven loops, these results reveal a simple all-order structure in the static sector and open a direct route to further higher orders.

\paragraph*{Conventions.}
The mostly minus metric, with signature $(+, -, -, -)$, is employed throughout this work, and multi-loop integrals are defined within the dimensional regularization scheme, considering $d=3+\epsilon$ spatial dimensions,
with $\epsilon$ being the analytic regulator. \\

\PRLsection{Gravitational dynamics in the static sector}
The conservative dynamics of a binary system of coalescing compact objects is characterized by a well-defined hierarchy of scales: the size of the compact objects $R_s$ is much smaller than the orbital separation $r$, which in turn is much smaller than the wavelength of the emitted gravitational radiation $\lambda$:
$
R_s \ll r \ll \lambda \, .
$
The first inequality defines a weak-coupling expansion in the ratio $R_s/r$, while the second corresponds to a small-velocity expansion, since $r/\lambda \sim v \ll 1$.

Within the post-Newtonian approximation, the expansion parameter is the typical velocity $v$ of the binary system, which is related to the gravitational potential through the virial theorem, $v^2 \sim G_N M/r$. Accordingly, one has
$
{G_N M}/{r} \sim v^2 \ll 1 \, ,
$
where $M = m_1 + m_2$ denotes the total mass of the system.\\

In this work, we focus on the static contributions to the gravitational potential, corresponding to terms of order ${\cal O}(G_{N}^{\pnord+1} v^0)$ at the $\pnord$-th PN order. We adopt the Effective Field Theory framework for the PN expansion, as introduced in~\cite{Goldberger:2004jt}, and analyze the all-order structure in terms of Feynman diagrams and correlation functions. \\

The bulk action governing the gravitational dynamics 
is given by the gauge-fixed Einstein-Hilbert action:
\begin{equation}
\label{eq:EH_GF}
\begin{split}
    S_{\rm bulk} &= S_{\rm EH} + S_{\rm GF} \\
    &= - \frac{ c^4}{16 \pi G_d} \int \dd^{d+1} x \sqrt{\vert g\vert} 
    \left( R - \frac{1}{2} g_{\mu\nu} \Gamma^\mu \Gamma^\nu \right)\, ,
\end{split}
\end{equation}
where $G_d=G_N(\sqrt{4 \pi\ e^{\gamma_E}} R_0)^{(d-3)}$ is the $d$-dimensional gravitational coupling, with
$\gamma_E$ the Euler-Mascheroni constant and $R_0$ an arbitrary length scale, with the gauge fixing term enforcing harmonic gauge ($\Gamma^\mu=g^{\nu\rho} \Gamma^{\mu}_{\nu\rho}=0$). 
The compact objects are described in terms of point-particle worldline actions
\begin{equation}
    S_{\rm pp}  = \sum_{a = 1, 2} \int \dd \tau \left( - m_a c \sqrt{g_{\mu\nu}u^\mu_{(a)}u^\nu_{(a)}} \right)  \, ,
    \label{eq:pp_action}
\end{equation}
where $u^\mu_{(a)} = {\rm d} x^\mu_{(a)}/{{\rm d}\tau}$ is the four-velocity. 
In the weak field approximation, we expand the metric around the Minkowski background, $g_{\mu\nu} = \eta_{\mu\nu} + h_{\mu\nu}$,
with $h_{\mu\nu}$ the graviton field.
The non-relativistic limit splits the graviton field into potential modes $H_{\mu\nu}$,
which are short-ranged and off-shell and scale as $(k_0,\textbf{k})\sim(v/r,1/r)$, 
and radiation modes $\bar{h}_{\mu\nu}$, 
which are long-range and on-shell and scale as $(k_0,\textbf{k})\sim(v/r,v/r)$.
In the following, we focus on the conservative static contributions to the potential, which are sourced by the potential modes, and therefore restrict attention to the $H_{\mu\nu}$ field.

We decompose the potential gravitons $H_{\mu\nu}$ using the Kaluza-Klein (KK) decomposition of the metric ($g_{\mu\nu} = \eta_{\mu\nu} + H_{\mu\nu}$)
introduced in~\cite{Kol:2007rx, Kol:2007bc} as:
\begin{eqnarray}
    g_{\mu\nu} = e^{2\bm{\phi} / c^2}
\begin{pmatrix}
 1 & - {\bm{A}_j / c^2}\\
- {\bm{A}_i / c^2} \,\,\,\,\,\, & -e^{-c_d\bm{\phi}/c^2}\bm{\gamma}_{ij}
+ {\bm{A}_i\bm{A}_j / c^4}  
\end{pmatrix}\, ,
\end{eqnarray}
where $c_d = 2(d-1)/(d-2)$ and $\bm{\gamma}_{ij} = \delta_{ij} + \bm{\sigma_{ij}}/c^2$. 
This parametrization rewrites the metric in terms of a scalar field $\bm{\phi}$,
a $d$-dimensional vector $\bm{A}_i$, and a $d$-dimensional symmetric tensor $\bm{\sigma}_{ij}$.
Substituting the parametrization into the bulk action~\eqref{eq:EH_GF}, and expanding it in the static limit, the $\bm{A}_i$ field appears only quadratically and therefore contributes only through quantum graviton loop corrections,
which are discarded in the classical limit.

The action in the static limit reduces to \cite{Kol:2010si}:
\begin{equation}
    \label{eq:EH_action_static}
    \begin{split}
            S_{\rm bulk}^{(v^0)} 
                = 
                - \frac{ c^4}{16 \pi G_d} \int \dd^{d+1} x 
                & \sqrt{\bm{\gamma}}  \left(\frac{c_d}{2 \,c^4}\,  \bm{\gamma^{ij} }\partial_i \bm{\phi} \partial_j \bm{\phi} \right. 
                \\
            &
            \left. - R\left[\bm{\gamma}\right] + \frac{1}{2} \big| \Gamma_i[\bm{\gamma}] \big|^2 \right) \, .
    \end{split}
\end{equation}
Analogously, substituting the KK variables in the point-particle action, and imposing the static limit $u^\mu_a = (1, \mathbf{0})$ we obtain:
\begin{equation}
    \label{eq:Spp}
    S_{\rm pp}^{(v^0)}  = \sum_{a = 1, 2} \int \dd t \left( - m_a c^2 e^{\bm{\phi}(x_{(a)})/c^2} \right)\, .
\end{equation}
where only the $\bm{\phi}$ field appears.

An effective action that depends only on the worldline parameters can be obtained by integrating out the potential gravitational
degrees of freedom of the theory. In the static limit, this reads as
\begin{equation}
\label{eq:Z_path_integral}
\begin{split}
        Z[\{\bm{x}^\mu_{(a)}\}] &=
        e^{\frac{\ii}{\hbar} \int \dd t\, (\mathcal{T} - \mathcal{V}_{\text{eff}})(\{\bm{x}^\mu_{(a)}\})}\\
        &= \frac{1}{Z_0}\int {\rm D}[\bm{\phi}, \bm{\sigma}_{ij}] e^{\frac{\ii}{\hbar} \left( S_{\rm pp}%\left(\{x^\mu_a\}, \phi, \sigma_{ij}\right) 
        + S_{\rm bulk}
        %\left(\phi, \sigma_{ij}\right) 
        \right)} \, ,
\end{split}
\end{equation}
with $S_{\rm pp} = S_{\rm pp}[\{\bm{x}_a^\mu\},\bm{\phi},\bm{\sigma}_{ij}]$, 
$S_{\rm bulk} = S_{\rm bulk}[\bm{\phi},\bm{\sigma}_{ij}]$, and
\begin{equation}
    Z_0 =  \int {\rm D}[\bm{\phi}, \bm{\sigma}_{ij}]\, e^{\ioverhbar S_{\rm bulk}}\, .
\end{equation}
In Eq.~\eqref{eq:Z_path_integral} $\mathcal{T}$ denotes the kinetic term density, while $\mathcal{V}_{\text{eff}}$ is the effective potential density. Throughout, we use calligraphic letters to denote quantities defined at the level of the time integrand.
The functional integration can be performed perturbatively,
expressing the effective action in terms of one-particle irreducible classical connected Feynman diagrams,
\begin{align} \label{eq:effective_potential_diagrammatic}
\mathcal{V}_{\text{eff}}= \ii \lim_{d\rightarrow 3} \int \frac{d^d \textbf{p}}{(2\pi)^d} ~e^{\ii \textbf{p}\cdot (\textbf{x}_{(1)}-\textbf{x}_{(2)})} \ \times  
\parbox{15mm}{
\begin{tikzpicture}[line width=0.7 pt, scale=0.25]
    \begin{scope}[shift={(0,0)}]
        \filldraw[color=gray!40, fill=gray!40, thick](0,0) rectangle (3,3);
        \draw (-1,0)--(4,0);
        \draw (-1,3)--(4,3);
    \end{scope}
\end{tikzpicture}
} \, . 
\end{align}
In the static sector the time dependence trivializes, hence here and in the remainder of this work we assume the overall time integral is understood. We also set $c=1$ for convenience. \\

\renewcommand{\arraystretch}{1.4}
\setlength{\tabcolsep}{2.5pt} % default is ~6pt
    \begin{table}[t]
    \centering
    \begin{tabular}{|c|c|c|c|c|c|c|c|}
    \hline 
    SF &  $0\times6$ & $2\times4$ & $0^3\times4$ & $0^2\times2^2$ & $0^5\times 2$ & ~$0^8~$ & Diagrams \\ \hline\hline
    0  & 28 & 5 & 5 & 1 & 1 & 1 & 41\\ 
    1 & 218 & 52 & 80 & 23 & 20 & 4 & 397 \\ 
    2  & 486 & 135 & 354 & 136 & 117 & 14 & 1242\\ 
    3  & 702 & 203 & 672 & 288 & 269 & 28 & 2162\\ 
    \hline\hline
    Total   & 1434 & 395 & 1111 & 448 & 407 & 47 & 3842\\
    \hline
    \end{tabular}
    \caption{
    Number of diagrams contributing to the 7PN static potential: displayed according to the self force contribution (rows) and 
    grouped in factorization classes (column) shown in \eqref{eq:V7PNclasses}, where number in the product denotes the diagrams at a PN order.
    }
    \label{tbl_no_diag}
\end{table}

\PRLsection{Factorization from Feynman graphs
}
As proved in Ref.~\cite{Foffa:2019hrb},
static diagrams at odd PN orders can be factorized in terms of products of lower PN order diagrams.
It is useful to distinguish between \textit{prime} and \textit{factorizable} static 
graphs. A prime graph (p.g.) contains only linear matter--$\bm{\phi}$ couplings of the form $m\bm{\phi}$, 
namely each $\bm{\phi}$ field emitted from the bulk couples individually to a worldline. 
Instead, a factorizable graph contains at least one composite worldline vertex $m\bm{\phi}^n$ with $n>1$, 
and can therefore be obtained by sewing together two or more lower-order subgraphs at that vertex (see Fig. 1 of Ref.~\cite{Foffa:2019hrb} for examples). 
Ref.~\cite{Foffa:2019hrb} proves that static prime graphs exist only at even PN orders; 
equivalently, all odd-PN static graphs are factorizable.
At 7PN, therefore, no new prime static graph appears.
 The full answer is obtained by combining lower-order even-PN building blocks. 
 As in the 5PN analysis of Ref.~\cite{Foffa:2019hrb}, the factorized contributions are organized according 
 to their sewing pattern. Importantly, each factorized graph is not given by a simple product of lower-order contributions: 
 when two subgraphs are sewn together, one must also include the factor associated with the new composite worldline vertex 
 and the appropriate combinatorial weight.
By the factorization theorem, the contribution of a factorizable graph ${\cal G}$ to the $\pnord$PN potential can be expressed in terms of the product of contributions from lower-PN order subgraphs ${\cal G}_i$ as:
\begin{align}
\cV_\pnord^{{\cal G}}
=
{\cal C}_{\pnord}\prod_{\pnord_i}
\frac{\cV_{\pnord_i}^{{\cal G}_i}}{{\cal C}_{\pnord_i}} \ ,
\label{eq:factorization_thm_std}
\end{align}
with $\sum_i \pnord_i=\pnord-1$,
 where ${\cal C}_{\pnord}$ and ${\cal C}_{\pnord_i}$ 
 are the symmetry factors of ${\cal G}$ and ${\cal G}_i$, respectively~as proposed in \cite{Foffa:2019hrb}. 
At the considered 7PN order, we remark that the determination of ${\cal C}_{\pnord}$ requires the generation of the complete set of 3842 contributing 7-loop diagrams computed using \texttt{PNTHR}: Post-Newtonian
Toolkit for Hamiltonian and Radiation~\cite{panther:inprep}.
Thus, owing to Eq.~\eqref{eq:factorization_thm_std},
the 7PN static contribution is obtained by summing all admissible sewings of lower-order even-PN prime sectors, with the appropriate ${\cal C}$ factors included.  
Accordingly, the 7PN correction to the effective potential is organized into six factorization classes, as
\begin{align}
\label{eq:V7PNclasses}
\cV_{7\mathrm{PN}}^{G_N^8}
&=
\cV_{0\mathrm{PN}\times 6\mathrm{PN}}
+
\cV_{2\mathrm{PN}\times 4\mathrm{PN}} 
+
\cV_{(0\mathrm{PN})^3\times 4\mathrm{PN}}
\nonumber\\
&
+
\cV_{(0\mathrm{PN})^2\times (2\mathrm{PN})^2}
+
\cV_{(0\mathrm{PN})^5\times 2\mathrm{PN}}
+ \cV_{(\mathrm{0PN})^8}
\, ,
\end{align}
and can be shown graphically as:
\begin{align}
\label{eq:V7PNclasses_graphix}
    \cV_{7\mathrm{PN}}^{G_N^8}
    &=
    % 0PNx6PN
    \left(\vcenter{\hbox{
        \begin{tikzpicture}[line width=0.8 pt, scale=0.25]
\draw[compobj] (-1,1.5)--(1,1.5);
\draw[compobj] (-1,-1.5)--(1,-1.5);
\draw[white,fill=white] (-0.5,-0.2) rectangle (-0.4,-0.4);
\draw[white,fill=white] (0.5,0.2) rectangle (0.4,0.4);
\draw[phigrav] (0,1.5)--(0,-1.5);
\node[blackdot] at (0,1.5) {};
\node[blackdot] at (0,-1.5) {};
        \end{tikzpicture}
    }}\right)
    \left(\vcenter{\hbox{
        \begin{tikzpicture}[line width=0.8 pt, scale=0.25]
        \draw[compobj] (-2,1.5)--(2,1.5);
\draw[compobj] (-2,-1.5)--(2,-1.5);
\draw[sigmagrav] (0,0.5)--(0,-0.5);
\draw[sigmagrav] (0,0.5)--(0.45,1);
\draw[sigmagrav] (0,0.5)--(1.3,0);
\draw[sigmagrav] (0,-0.5)--(-0.45,-1);
\draw[sigmagrav] (0,-0.5)--(-1.3,0);
\draw[white,fill=white] (-0.5,-0.2) rectangle (-0.4,-0.4);
\draw[white,fill=white] (0.5,0.2) rectangle (0.4,0.4);
\draw[phigrav] (-1.3,1.5)--(-1.3,-1.5);
\draw[phigrav] (0.45,1.5)--(0.45,-1.5);
\draw[phigrav] (-0.45,1.5)--(-0.45,-1.5);
\draw[phigrav] (1.3,1.5)--(1.3,-1.5);
\node[blackdot] at (0,0.5) {};
\node[blackdot] at (0,-0.5) {};
\node[blackdot] at (0.45,1) {};
\node[blackdot] at (1.3,0) {};
\node[blackdot] at (-0.45,-1) {};
\node[blackdot] at (-1.3,0) {};
\node[blackdot] at (-1.3,1.5) {};
\node[blackdot] at (-1.3,-1.5) {};
\node[blackdot] at (0.45,1.5) {};
\node[blackdot] at (0.45,-1.5) {};
\node[blackdot] at (-0.45,1.5) {};
\node[blackdot] at (-0.45,-1.5) {};
\node[blackdot] at (1.3,1.5) {};
\node[blackdot] at (1.3,-1.5) {};
        \end{tikzpicture}
    }} 
    + 333 \ {\rm p.g.} \right)
    \nonumber \\   & 
    +
    % 2PNx4PN
    \left(\vcenter{\hbox{
        \begin{tikzpicture}[line width=0.8 pt, scale=0.25]
\draw[compobj] (-1,1.5)--(1,1.5);
\draw[compobj] (-1,-1.5)--(1,-1.5);
\draw[white,fill=white] (-0.5,-0.2) rectangle (-0.4,-0.4);
\draw[white,fill=white] (0.5,0.2) rectangle (0.4,0.4);
\draw[phigrav] (-0.425,1.5)--(-0.425,-1.5);
\draw[phigrav] (0.425,1.5)--(0.425,-1.5);
\draw[sigmagrav] (-0.425,0)--(0.425,0);
\node[blackdot] at (-0.425,0) {};
\node[blackdot] at (0.425,0) {};
\node[blackdot] at (-0.425,1.5) {};
\node[blackdot] at (-0.425,-1.5) {};
\node[blackdot] at (0.425,1.5) {};
\node[blackdot] at (0.425,-1.5) {};
        \end{tikzpicture}
    }}
    + 3 \ {\rm p.g.} \right)
    \left(\vcenter{\hbox{
        \begin{tikzpicture}[line width=0.8 pt, scale=0.25]
    \draw[compobj] (-1.5,1.5)--(1.5,1.5);
    \draw[compobj] (-1.5,-1.5)--(1.5,-1.5);
    \draw[sigmagrav] (-0.85,0)--(-0.325,0.8);
    \draw[sigmagrav] (-0.325,0.8)--(0.85,0);
    \draw[white,fill=white] (0.05,0.7) rectangle (-0.05,0.05);
    \draw[sigmagrav] (-0.325,0.8)--(0,-0.1);
    \draw[phigrav] (-0.85,1.5)--(-0.85,-1.5);
    \draw[phigrav] (0,1.5)--(0,-1.5);
    \draw[phigrav] (0.85,1.5)--(0.85,-1.5);
\node[blackdot] at (-0.85,1.5) {};
\node[blackdot] at (0,-0.1) {};
\node[blackdot] at (-0.85,-1.5) {};
\node[blackdot] at (0,1.5) {};
\node[blackdot] at (0,-1.5) {};
\node[blackdot] at (-0.325,0.8) {};
\node[blackdot] at (0.85,1.5) {};
\node[blackdot] at (0.85,-1.5) {};
\node[blackdot] at (-0.85,0) {};
\node[blackdot] at (0.85,0) {};
        \end{tikzpicture}
    }} 
    + 42 \ {\rm p.g.} 
    \right) 
    \nonumber \\ & 
    +
    % 0PN^3x4PN
    \left(\vcenter{\hbox{
        \begin{tikzpicture}[line width=0.8 pt, scale=0.25]
\draw[compobj] (-1,1.5)--(1,1.5);
\draw[compobj] (-1,-1.5)--(1,-1.5);
\draw[white,fill=white] (-0.5,-0.2) rectangle (-0.4,-0.4);
\draw[white,fill=white] (0.5,0.2) rectangle (0.4,0.4);
\draw[phigrav] (0,1.5)--(0,-1.5);
\node[blackdot] at (0,1.5) {};
\node[blackdot] at (0,-1.5) {};
        \end{tikzpicture}
    }}\right)^3
    \left(\vcenter{\hbox{
        \begin{tikzpicture}[line width=0.8 pt, scale=0.25]
    \draw[compobj] (-1.5,1.5)--(1.5,1.5);
    \draw[compobj] (-1.5,-1.5)--(1.5,-1.5);
    \draw[sigmagrav] (-0.85,0)--(-0.325,0.8);
    \draw[sigmagrav] (-0.325,0.8)--(0.85,0);
    \draw[white,fill=white] (0.05,0.7) rectangle (-0.05,0.05);
    \draw[sigmagrav] (-0.325,0.8)--(0,-0.1);
    \draw[phigrav] (-0.85,1.5)--(-0.85,-1.5);
    \draw[phigrav] (0,1.5)--(0,-1.5);
    \draw[phigrav] (0.85,1.5)--(0.85,-1.5);
\node[blackdot] at (-0.85,1.5) {};
\node[blackdot] at (0,-0.1) {};
\node[blackdot] at (-0.85,-1.5) {};
\node[blackdot] at (0,1.5) {};
\node[blackdot] at (0,-1.5) {};
\node[blackdot] at (-0.325,0.8) {};
\node[blackdot] at (0.85,1.5) {};
\node[blackdot] at (0.85,-1.5) {};
\node[blackdot] at (-0.85,0) {};
\node[blackdot] at (0.85,0) {};
        \end{tikzpicture}
    }} 
    + 42 \ {\rm p.g.} \right)
    \nonumber \\   & 
    +
    % 0PN^2x2PN^2
    \left(\vcenter{\hbox{
        \begin{tikzpicture}[line width=0.8 pt, scale=0.25]
\draw[compobj] (-1,1.5)--(1,1.5);
\draw[compobj] (-1,-1.5)--(1,-1.5);
\draw[white,fill=white] (-0.5,-0.2) rectangle (-0.4,-0.4);
\draw[white,fill=white] (0.5,0.2) rectangle (0.4,0.4);
\draw[phigrav] (0,1.5)--(0,-1.5);
\node[blackdot] at (0,1.5) {};
\node[blackdot] at (0,-1.5) {};
        \end{tikzpicture}
    }}\right)^2
    \left(\vcenter{\hbox{
        \begin{tikzpicture}[line width=0.8 pt, scale=0.25]
\draw[compobj] (-1,1.5)--(1,1.5);
\draw[compobj] (-1,-1.5)--(1,-1.5);
\draw[white,fill=white] (-0.5,-0.2) rectangle (-0.4,-0.4);
\draw[white,fill=white] (0.5,0.2) rectangle (0.4,0.4);
\draw[phigrav] (-0.425,1.5)--(-0.425,-1.5);
\draw[phigrav] (0.425,1.5)--(0.425,-1.5);
\draw[sigmagrav] (-0.425,0)--(0.425,0);
\node[blackdot] at (-0.425,0) {};
\node[blackdot] at (0.425,0) {};
\node[blackdot] at (-0.425,1.5) {};
\node[blackdot] at (-0.425,-1.5) {};
\node[blackdot] at (0.425,1.5) {};
\node[blackdot] at (0.425,-1.5) {};
        \end{tikzpicture}
    }}
    + 3 \ {\rm p.g.} \right)^2
    \nonumber \\   & 
+
    % 0PN^5x2PN
    \left(\vcenter{\hbox{
        \begin{tikzpicture}[line width=0.8 pt, scale=0.25]
\draw[compobj] (-1,1.5)--(1,1.5);
\draw[compobj] (-1,-1.5)--(1,-1.5);
\draw[white,fill=white] (-0.5,-0.2) rectangle (-0.4,-0.4);
\draw[white,fill=white] (0.5,0.2) rectangle (0.4,0.4);
\draw[phigrav] (0,1.5)--(0,-1.5);
\node[blackdot] at (0,1.5) {};
\node[blackdot] at (0,-1.5) {};
        \end{tikzpicture}
    }}\right)^5 
%    \! \times \!
    %
    \left(\vcenter{\hbox{
        \begin{tikzpicture}[line width=0.8 pt, scale=0.25]
\draw[compobj] (-1,1.5)--(1,1.5);
\draw[compobj] (-1,-1.5)--(1,-1.5);
\draw[white,fill=white] (-0.5,-0.2) rectangle (-0.4,-0.4);
\draw[white,fill=white] (0.5,0.2) rectangle (0.4,0.4);
\draw[phigrav] (-0.425,1.5)--(-0.425,-1.5);
\draw[phigrav] (0.425,1.5)--(0.425,-1.5);
\draw[sigmagrav] (-0.425,0)--(0.425,0);
\node[blackdot] at (-0.425,0) {};
\node[blackdot] at (0.425,0) {};
\node[blackdot] at (-0.425,1.5) {};
\node[blackdot] at (-0.425,-1.5) {};
\node[blackdot] at (0.425,1.5) {};
\node[blackdot] at (0.425,-1.5) {};
%\node at (0,2.2) {\scalebox{0.9}{(3SF)}};	
%\node at (0,-2.4) {\scalebox{0.9}{+3}};
        \end{tikzpicture}
    }} 
    +3 \ {\rm p.g.} \right)
    %
    % \nonumber \\   & 
    % %
    +
    % 0PN^8
    \left(\vcenter{\hbox{
        \begin{tikzpicture}[line width=0.8 pt, scale=0.25]
\draw[compobj] (-1,1.5)--(1,1.5);
\draw[compobj] (-1,-1.5)--(1,-1.5);
\draw[white,fill=white] (-0.5,-0.2) rectangle (-0.4,-0.4);
\draw[white,fill=white] (0.5,0.2) rectangle (0.4,0.4);
\draw[phigrav] (0,1.5)--(0,-1.5);
\node[blackdot] at (0,1.5) {};
\node[blackdot] at (0,-1.5) {};
%\node at (0,2.2) {\scalebox{0.9}{(3SF)}};	
%\node at (0,-2.4) {\scalebox{0.9}{+0}};
        \end{tikzpicture}      
    }}\right)^8
%    \nonumber \\ & 
    %
    \, ,
\end{align}
where the black, blue and green lines 
represent the massive bodies, the $\phi$ and 
$\sigma$ fields, respectively.
The number of diagrams in each factorized class\footnote{
For instance, the first class of factorised diagrams in Eq.\eqref{eq:V7PNclasses}, 
{\it i.e.} $\cV_{0\mathrm{PN}\times 6\mathrm{PN}}$,  
graphically represented by the first term in the {\it r.h.s.} of Eq.\eqref{eq:V7PNclasses_graphix}, 
takes contribution from 1434 graphs (as reported in the bottom entry of the first column of Table~\ref{tbl_no_diag}); 
while the last class of factorised diagrams in Eq.\eqref{eq:V7PNclasses},
{\it i.e.} $\cV_{(0\mathrm{PN})^8}$,
represented by the last term in the {\it r.h.s.} of Eq.\eqref{eq:V7PNclasses_graphix}, 
takes contribution from 47 graphs (as reported in the bottom entry of the last column of Table~\ref{tbl_no_diag}).
}  
is reported in Table~\ref{tbl_no_diag}, and grouped according to their $n^{\rm th}$-order self force ($n$SF) contribution which corresponds to the terms proportional to $m_1^{8-n} m_2^{1+n}$.

The prime graphs appearing on the right-hand side of
Eqs.~(\ref{eq:V7PNclasses}, \ref{eq:V7PNclasses_graphix})
reduce to Euclidean multi-loop massless two-point integrals
in \(d=3+\epsilon\) dimensions~\cite{Foffa:2016rgu}, with even
loop order \(\ell=\{0,2,4,6\}\). The topologies required up to
\(V_{6{\rm PN}}\) were already computed in
Ref.~\cite{Brunello:2025gpf}, so the present calculation can be
carried out within the same computational framework. The corresponding integration-by-parts (IBP) reductions 
\cite{Tkachov:1981wb,Chetyrkin:1981qh,Laporta:2001dd}
are performed analytically with
the code \texttt{PRISM}~\cite{PRISM}, combining spanning
cuts~\cite{Larsen:2015ped,Lee:2013mka,Maierhofer:2017gsa},
syzygy-based
methods~\cite{Gluza:2010ws,Wu:2023upw,Wu:2025aeg,Smith:2025xes}
implemented with \texttt{Singular}~\cite{DGPS}, improved
seeding
algorithms~\cite{Lange:2025fba,Wu:2023upw,Wu:2025aeg}, and
finite-field reconstruction through
\texttt{FiniteFlow}~\cite{Peraro:2019svx}. Analytic expressions
for the relevant master integrals at 4PN and 6PN can be found, respectively, in Ref.~\cite{Foffa:2016rgu} and in Ref.~\cite{Brunello:2025gpf}. \\

\PRLsection{Factorization from correlation functions 
}
Building on a proposed computational framework based on correlation functions~\cite{correlators2026}, we derive the effective static potential in terms of connected gravitational correlators. We show that the factorization theorem is a consequence of the ${\mathbb Z}_2$ symmetry of the static sector.

\PRLsubsection{a) Correlation functions framework}
The effective action of the system can be evaluated as the logarithm of the full path integral in Eq.~\eqref{eq:Z_path_integral}.
The latter can be recast as the bulk expectation value of the operator $\mathcal{O} = e^{\ioverhbar \mathcal{S}_{\rm pp}}$,
\begin{equation}
\label{eq:S_eff_expectation_value}
\begin{split}
       % \int \dd t\, 
       \left(\mathcal{T}-\mathcal{V}_{\text{eff}}[\{\bm{x}^\mu_{(a)}\}]\right) = - \ii \hbar \log\left( \langle  e^{\ioverhbar \mathcal{S}_{\rm pp}} \rangle_{\rm bulk} \right)  \, ,
\end{split}
\end{equation}
where we define the bulk expectation value of a generic operator $\mathcal{O}[\bm{\phi}, \bm{\sigma}_{ij} ]$ as the path integral over the gravitational fields, 
\begin{equation}
\!\!\!
    \langle \mathcal{O}[\bm{\phi}, \bm{\sigma}_{ij} ] \rangle_{\rm bulk} = \frac{1}{Z_0} \int {\rm D}[\bm{\phi}, \bm{\sigma}_{ij}]\, \mathcal{O}[ \bm{\phi}, \bm{\sigma}_{ij} ]\, e^{\ioverhbar \mathcal{S}_{\rm bulk}}\, .
\end{equation}
In the static sector, the time dependence drops out, and the path integral is evaluated over fields on a $d$-dimensional Euclidean space.
We note that the point-particle action of Eq.~\eqref{eq:Spp} localizes the fields $\bm{\phi}(x)$ to be evaluated either at the worldline position $\bm{x}_{(1)}$ or $\bm{x}_{(2)}$. Hence we define $\bm{\phi}_a = \bm{\phi}(\bm{x}_{(a)})$ and the composite operators
$
%\begin{equation}
%    \label{eq:composite_operators}
    \Phi_1 := e^{\bm{\phi}_1}\ , \ 
    \Phi_2 := e^{\bm{\phi}_2} \, ,
%\end{equation}
$
so that the point-particle action reads
$
%\begin{equation}
%    \label{eq:Spp_insertions}
    \mathcal{S}_{\rm pp} = - \left( m_1 \Phi_1 + m_2  \Phi_2 \right)\, .
%\end{equation}
$
We evaluate the bulk expectation value in Eq.~\eqref{eq:S_eff_expectation_value} perturbatively, performing a series expansion to recover a polynomial expression in the gravitational fields:
\begin{equation}
    \label{eq:Z_expansion}
    \!\!\!
    \Big\langle e^{\,\frac{\ii}{\hbar}\mathcal{S}_{\rm pp}}\Big\rangle_{\rm bulk}
    =
    \sum_{a,b\ge 0} \left(- \frac{\ii}{\hbar}\right)^{a+b}
    \frac{m_1^a}{a!}\frac{m_2^b}{b!}\,
    \langle \Phi_1^a \Phi_2^b\rangle_{\rm bulk}
    \, .
\end{equation}

The expectation value of $\langle \Phi_1^a \Phi_2^b\rangle_{\rm bulk}$ can be further expanded perturbatively. By linearity of the expectation value, the effective static post-Newtonian action is then entirely expressed in terms of gravitational correlation functions of the $\bm{\phi}$ field. 

We introduce the connected correlation functions $\Gamma_{n_1,n_2}$ of the $\bm{\phi}$ field, evaluated at the worldline positions,
\begin{equation}
\label{eq:Gamma_n1n2_definition}
    \Gamma_{n_1,n_2} = \left(\ioverhbar\right)^{n_1 + n_2 - 1} \big\langle \bm{\phi}_1^{n_1} \bm{\phi}_2^{n_2} \big\rangle_c = 
    \vcenter{\hbox{
        \begin{tikzpicture}[line width=0.8 pt, scale=0.22]
            
            %\draw[color=gray, dashed] (-2, 4) -- (5, 4);
            %\draw[color=gray, dashed] (-2, 0) -- (5, 0);
            
            \draw[color=blue] (1.5, 2) -- (-0.5, 4);
            \draw[color=blue] (1.5, 2) -- (3.5, 4);
            
            \node[font=\scriptsize] at (1.5, 4) {$\cdot\!\cdot\!\cdot\, n_1 \!\cdot\!\cdot \cdot$};
        
            \draw[color=blue] (1.5, 2) -- (-0.5, 0);
            \draw[color=blue] (1.5, 2) -- (3.5, 0);
            \node[font=\scriptsize] at (1.5, 0) {$\cdot\!\cdot\!\cdot\, n_2 \!\cdot\!\cdot \cdot$};

            % bullets at endpoints
            \fill[color=black] (-0.5, 4) circle (0.2);
            \fill[color=black] (3.5, 4) circle (0.2);
            \fill[color=black] (-0.5, 0) circle (0.2);
            \fill[color=black] (3.5, 0) circle (0.2);
        
            \filldraw[color=gray!40, fill=gray!40, thick] (1.5, 2) circle (1.25);
            
            %\filldraw[pattern=north east lines] (1.5, 2) circle (1.1);
            
        \end{tikzpicture}
    }} \, ,
\end{equation}
normalized such that its leading order classical scaling is the real quantity
$
%\begin{equation}
    \Gamma_{n_1, n_2}\propto \hbar^0 %\left(\frac{G_N\,  } {r}\right)
    (G_N/r)
    ^{n_1+n_2-1}  \, .
%\end{equation}
$
The connected correlation functions are diagrammatically represented as sums over sets of prime graphs, i.e. the sum of connected Feynman diagrams with $n_1$ and $n_2$ $\bm{\phi}$ legs sourced from the worldlines via linear operators at positions $\bm{x}_{(1)}$ and $\bm{x}_{(2)}$, respectively. 

To evaluate the quantity $\langle \Phi_1^a \Phi_2^b\rangle_{\rm bulk}$ we introduce the linear sources $J_1$ and $J_2$ and the generating functional of the bulk connected correlation functions 
\begin{align}
\label{eq:W_bulk_Gamma_relation}
         W_{\rm bulk}(J_1,J_2)
         & = 
         -\ii \hbar \log \left\langle \exp\left[ \frac{\ii}{\hbar} \Big( J_1 \bm{\phi}_1 + J_2 \bm{\phi}_2 \Big) \right] \right\rangle_{\rm bulk} \nonumber
         \\
         & =  
         \sum_{n_1, n_2 \geq 1} \frac{J_1^{n_1} J_2^{n_2}}{n_1! n_2!} \Gamma_{n_1, n_2} \, ,
\end{align}
which have been directly collapsed to the localized worldlines positions. The sum in $W_{\rm bulk}$ is restricted to $n_a > 0$ to remove self-energies and vanishing 1-point functions. 
Using the identities 
$\langle O[\phi]\rangle = O[-\ii \hbar\frac{\delta}{\delta J}] Z[J] \bigr|_{J=0}$
and $\exp\left(\alpha \frac{\partial}{\partial J}\right) Z(J) = Z(J + \alpha)$, with $Z = \exp(\ii/\hbar \, W_{\rm bulk}[J])$~\cite{Srednicki_2007,kleinert2016particles}, it holds:
\begin{equation}
\label{eq:Phi_expectation_value}
\begin{split}
      \langle \Phi_1^a \Phi_2^b\rangle_{\rm bulk} &= \exp\left(\ioverhbar W_{\rm bulk}
     \left(
      - \ii \hbar \, a,
      - \ii \hbar\, b
      \right) \right) \, .
\end{split}
\end{equation}
The expression of the bulk expectation value in Eq.~\eqref{eq:Z_expansion} then 
can be expressed entirely in terms of the connected gravitational correlation functions $\Gamma_{n_1,n_2}$.

\PRLsubsection{b) $\mathbb{Z}_2$ symmetry of the static sector}
Employing the KK parametrization of the metric~\cite{Kol:2007rx, Kol:2007bc}, the classical $d$-dimensional static bulk action of Eq.~\eqref{eq:EH_action_static} is manifestly invariant under a $\mathbb{Z}_2$ symmetry of the $\bm{\phi}$ gravitational field~\cite{Kol:2011vg}, 
i.e., the action is invariant under the discrete transformation
$
%\begin{equation}
    \bm{\phi} \;\rightarrow\; -\bm{\phi} \, .
%\end{equation}
$
This $\mathbb{Z}_2$ symmetry is a subgroup of $\text{SL}(2,\mathbb{R})$ symmetry of the static Einstein-Hilbert action~\cite{Ehlers:1959aug, Buchdahl:1959nk, harrison1968new, Ernst:1967wx, Geroch:1970nt, stephani2009exact,Kol:2011vg,Parra-Martinez:2025bcu} in the $d=3$ spatial dimensions.

We assume the Minkowski vacuum to preserve the symmetry, $\langle \bm{\phi} \rangle = 0$.
Then the $\mathbb{Z}_2$ symmetry implies selection rules for the correlation functions. Specifically, only correlation functions with an even total number of $\bm{\phi}$ fields can be non-vanishing: 
%
%$
\begin{equation}
    \label{eq:Z2_selection_rule}
     \Gamma_{n_1,n_2}=0 \qquad \text{if } n_1+n_2 \in 2\mathbb{N} + 1\, .
\end{equation}
%$
This accidental symmetry of the bulk gravitational action, which also underlies the vanishing of black hole tidal Love numbers~\cite{Kol:2011vg,Parra-Martinez:2025bcu}, strictly constrains the static potential for binary black holes to all post-Newtonian orders.

\PRLsubsection{c) All-order structure of the static potential}
The all-order effective action can be obtained from Eqs.~\eqref{eq:S_eff_expectation_value}, \eqref{eq:Z_expansion} and \eqref{eq:Phi_expectation_value}, enforcing the $\mathbb{Z}_2$ symmetry:% via 

    \begin{align}
     & \mathcal{S}_{\rm eff} =
     \mathcal{T} -    \cV_{\rm eff}
     = - \ii \hbar 
     \log\left[  
     \sum_{a,b\geq0} \left(-\frac{\ii}{\hbar}\right)^{a+b} 
     \frac{m_1^a}{a!} 
     \frac{m_2^b}{b!} 
     \quad \ 
     \right.
      \nonumber \\ &
     \left.
     \!\!\!
     \times 
     \exp\left( \sum_{\substack{n_1,n_2\ge 1\\ n_1+n_2 \ \mathrm{even}}} \left(-\ii \hbar\right)^{n_1+n_2 - 1} \frac{a^{n_1}}{n_1!} \frac{b^{n_2}}{n_2!} \Gamma_{n_1, n_2} \right) 
     \right] \!\!.
     \end{align}

This formula represents the first main
result of this Letter. It establishes the exact all-order structure of the static post-Newtonian effective action as a decomposition into fundamental {\it connected correlation functions} $\Gamma_{n_1,n_2}$.
The classical static potential at $\pnord$-PN order, $\mathcal{V}_{\pnord\rm{PN}}$ at ${\cal O}(G_N^{\pnord+1})$ is a mass polynomial of order $\mathcal{O}(m_i^{k+2}$) (for $i=1,2$) \cite{Damour:2019lcq}, which can be obtained by taking a series expansion in the masses 
and then evaluating the $\hbar \rightarrow 0$ limit:
\begin{equation}
\begin{split}
    \label{eq_all_order_V}
        % \cV_{\rm classical} & = \sum_{k=0}^{\infty}\mathcal{V}_{\pnord\rm{PN}}^{G_N^{\pnord+1}}\, ,     
        % \\ 
        \mathcal{V}_{\pnord\rm{PN}}^{G_N^{\pnord+1}} & = \sum_{l = 1}^{\pnord+1} m_1^l m_2^{\pnord-l+2} \frac{\widehat{\cV}_{l, \, \pnord-l+2}}{l! (\pnord-l+2)!} \, ,\\ 
        \widehat{\cV}_{r, s} & = -\lim_{\hbar\to  0}\, \frac{\partial^{r+s}}{\partial m_1^r 
         \, \partial m_2^{s}} \left(\mathcal{S}_{\rm eff} \right)
         \biggr|_{m_1=0, m_2=0}
%         {\substack{m_1 = 0 \\ m_2 = 0}}
         \, .
\end{split}
\end{equation}

The expressions of 
$\mathcal{V}_{\pnord\rm{PN}}^{G_N^{\pnord+1}}$,
for $\pnord \le 7$, in terms of $\Gamma_{n_1,n_2}$ are reported in App.~\ref{app:potential_correlators}.
A ready-to-use implementation of the formula is provided in the ancillary
file.

\PRLsubsection{d) Factorization theorem from $\mathbb{Z}_2$ symmetry}
The factorization theorem 
admits a reformulation in terms of correlation functions, based on Eq.~\eqref{eq_all_order_V} and its underlying $\mathbb{Z}_2$ symmetry. Each coefficient $\widehat{\cV}$ in Eq.~\eqref{eq_all_order_V} can be written as
$
%\begin{equation}
%\label{eq:factorization}
\widehat{\cV}_{l, \pnord-l+2}
= c_{l, \pnord-l+2}\,\Gamma_{l,\pnord-l+2}
+ \sum_{\alpha} c_\alpha \prod_{i=1}^{N_\alpha}
\Gamma_{n_{1,i}^{(\alpha)},\,n_{2,i}^{(\alpha)}} \, ,
%\end{equation}
$
with $\sum_{i = 1}^{N_\alpha} (
n_{1,i}^{(\alpha)}
+n_{2,i}^{(\alpha)}
)= \pnord+1+N_{\alpha}$, where $N_\alpha$ is the number of factorized correlators. 
Here, the correlator $\Gamma_{l,\pnord-l+2}$ appears linearly and corresponds to {\it prime} graphs, while the products of lower-order correlators $\Gamma_{n_{1,i}^{(\alpha)},\,n_{2,i}^{(\alpha)}}$ encode {\it factorizable} graphs generated by the composite worldline operators. % in Eq.~\eqref{eq:composite_operators}.
For odd $\pnord$, the prime correlators $\Gamma_{l,\pnord-l+2}$ vanish identically due to the $\mathbb{Z}_2$ selection rule in Eq.~\eqref{eq:Z2_selection_rule}. As a result, each $\widehat{\cV}_{l,\pnord-l+2}$, and thus the full classical odd-PN static potential $\widehat{\cV}_{\pnord{\rm PN}}^{G_N^{\pnord+1}}$, depends exclusively on products of correlators appearing at lower even orders.
This identifies the $\mathbb{Z}_2$ symmetry as the underlying origin of the factorization theorem of Ref.~\cite{Foffa:2019hrb}, and explains why odd-PN static contributions are fully determined by lower-order data.

\PRLsubsection{e) Known correlation functions}
Table~\ref{tab:gamma_values} reports the known nonzero correlators, obtained by matching Eq.~\eqref{eq_all_order_V} to the known static potentials through 6PN~\cite{Foffa:2016rgu,Foffa:2019hrb,Blumlein:2019zku,Brunello:2025gpf}.
\begin{table}[t]
%\vspace{0.5cm}
\renewcommand{\arraystretch}{1.4}
% \begin{ruledtabular}
\begin{tabular}{|c|c|c|c|c|}
\hline
PN& 0SF & 1SF & 2SF & 3SF 
\\
\colrule
0 & $\bar{\Gamma}_{1,1} = 1$ & & & 
\\
2 & $\bar{\Gamma}_{1,3} = 2$ & $\bar{\Gamma}_{2,2} = 8$ & & 
\\
4 & $\bar{\Gamma}_{1,5} = 24$ & $\bar{\Gamma}_{2,4} = 160$ & $\bar{\Gamma}_{3,3} = 504$ & 
\\
6 & $\bar{\Gamma}_{1,7} = 720$ & $\bar{\Gamma}_{2,6} = 6400$ & $\bar{\Gamma}_{3,5} = 31760$ & $\bar{\Gamma}_{4,4} = 83456$ 
\\
\hline
\end{tabular}
% \end{ruledtabular}
\caption{\label{tab:gamma_values} 
Rescaled $\bar{\Gamma}_{n_1,n_2}$ correlators in $d=3$ spatial dimensions. They are related to $\Gamma_{n_1, n_2}$ via $\Gamma_{n_1 , n_2} = \left(G_N/r\right)^{n_1 + n_2 - 1} \bar{\Gamma}_{n_1,n_2}$,
with
$\bar{\Gamma}_{n_1,n_2} = \bar{\Gamma}_{n_2,n_1}$. 
}
\end{table}
Notably, all correlators known thus far are positive integers that increase with PN order.

\PRLsubsection{f) All-order value of 0SF correlators} 
The 0SF correlators $\Gamma_{1,n} = \Gamma_{n,1}$ are fixed to all orders by matching Eq.~\eqref{eq_all_order_V} to the known test-particle limit, 
$
    \mathcal{V}^{m_1\ll m_2} = m_1  \sqrt{g_{00}} = 
    m_1 
\sqrt{(1-x)/(1+x)}
    \ ,
$
with $x=G_N m_2/r$,
yielding:
\begin{eqnarray}
\label{eq:0SF_correlator}
    \Gamma_{1,n}=  \vcenter{\hbox{
        \begin{tikzpicture}[line width=0.8 pt, scale=0.22]
            
            %\draw[color=gray, dashed] (-2, 4) -- (5, 4);
            %\draw[color=gray, dashed] (-2, 0) -- (5, 0);
            
            %\draw[color=blue] (1.5, 2) -- (-0.5, 4);
            %\draw[color=blue] (1.5, 2) -- (3.5, 4);
            \draw[color=blue] (1.5, 2) -- (1.5, 4);
            
            %\node[font=\scriptsize] at (1.5, 4) {$\cdot\!\cdot\!\cdot\, n_1 \!\cdot\!\cdot \cdot$};
        
            \draw[color=blue] (1.5, 2) -- (-0.5, 0);
            \draw[color=blue] (1.5, 2) -- (3.5, 0);
            \node[font=\scriptsize] at (1.5, 0) {$\cdot\!\cdot\!\cdot\, n \!\cdot\!\cdot \cdot$};

            % bullets at endpoints
            \fill[color=black] (1.5, 4) circle (0.2);
            %\fill[color=black] (3.5, 4) circle (0.2);
            \fill[color=black] (-0.5, 0) circle (0.2);
            \fill[color=black] (3.5, 0) circle (0.2);
        
            \filldraw[color=gray!40, fill=gray!40, thick] (1.5, 1.85) circle (1.25);
            
            %\filldraw[pattern=north east lines] (1.5, 2) circle (1.1);
            
        \end{tikzpicture}
    }}=  \left(\frac{G_N}{r}\right)^{n}(n-1)!  \quad \text{for }n\text{ odd}\, .
\end{eqnarray}
The derivation of this formula is presented in App.~\ref{app:test_body}.

\PRLsection{7PN static potential
%\label{sec:7PN_and_beyond}
}
The 7PN static contribution 
${\cal O}(G_N^8)$
of the two-body gravitational potential receives contributions from 
3842 7-loop PN-EFT diagrams (up to $m_1 \leftrightarrow m_2$ exchange).
Nevertheless, owing to 
the factorization theorem, 
it can be determined 
either using Eqs.~(\ref{eq:V7PNclasses}, 
\ref{eq:V7PNclasses_graphix}), 
in terms 
of up to 6-loop diagrams, 
or, more directly, via correlation functions using Eq.~\eqref{eq_all_order_V} and Table~\ref{tab:gamma_values}, both 
yielding:

\begin{align}
    \label{eq:7PN_result}
\mathcal{V}_{\rm 7PN}^{G_N^8} & = 
\frac{G_N^8}{r^8} 
\left(\frac{35}{128} m_1 m_2^8 
+\frac{248}{9} m_1^2 m_2^7 
+\frac{1059}{2} m_1^3 m_2^6 
\right.
\nonumber \\ &
\left.
\qquad \qquad
+\frac{89383}{36} m_1^4 m_2^5 
+ (1 \leftrightarrow 2) \right)\, .
\end{align}

This remarkably compact formula represents the second main result of this Letter. 
The coefficient of $m_1m_2^8$ agrees with the expansion of the test-particle
limit above, providing a direct check of Eq.~\eqref{eq:7PN_result}.
\\

\PRLsection{Beyond 7PN}
The master formula in Eq.~\eqref{eq_all_order_V}
can be used to derive the static contributions to higher PN orders in terms of correlation functions.
The 8PN and 9PN static potentials depend on four currently undetermined correlators, 
which require the evaluation of 8-loop integrals.
In general, at even $(2k)$-PN order, $k+1$ new prime correlation functions appear, reducing to $k$ via Eq.~\eqref{eq:0SF_correlator}.
Genuinely new prime correlators enter only at even PN orders, while odd-PN static potentials are fixed recursively by lower-order correlators. 

The expressions of 
the 8PN and 9PN static potentials are reported in App.~\ref{app:8_9_PN}. \\

\PRLsection{Conclusion}
We investigated the all-order structure of the post-Newtonian static gravitational potential within a field-theoretic framework. Exploiting the $\mathbb{Z}_2$ symmetry of the static sector, we derived a closed formula for the generic contribution at $n$PN order in terms of correlation functions. We showed that any odd-order term is entirely determined by lower-order data, thereby providing a novel theoretical formulation of the factorization theorem~\cite{Foffa:2019hrb}.  

Our results yield both conceptual and computational advantages. In its original formulation, the factorization theorem allows one to determine the odd $n$-th order contribution by avoiding the explicit evaluation of the corresponding $n$-loop diagrams, reducing the problem to lower-order (up to $(n-1)$-loop) computations. On the other hand, our closed formula, expressing odd-order contributions directly and solely in terms of known lower-order correlators (which are intrinsically sums of diagrams), 
eliminates the need for diagrammatic input altogether.

As a concrete application, we have computed for the first time the static gravitational potential at $7$PN order, and explicitly verified the consistency between our closed-form approach and the diagrammatic implementation of the factorization theorem up to this order.

More generally, the formalism developed here provides an efficient and systematic organization of static PN corrections at arbitrarily high orders. Once the lower-order connected correlators are known, all odd-PN contributions follow recursively, while genuinely new information enters only through even-PN correlators. This structure imposes strong constraints on higher-order terms and clarifies the origin of the remarkable predictive power observed in the static sector.
The correlator decomposition also suggests the possibility of resumming the static sector, as exemplified by Eq.~\eqref{eq:0SF_correlator}, which resums the test-particle limit.

Our findings further suggest that the static post-Newtonian expansion is governed by an underlying algebraic structure, in which symmetry and correlation functions organize the 
multi-loop, yet classical, two-body dynamics in a highly constrained and unexpectedly compact form.

\PRLsection{Acknowledgments}
We thank Mao Zeng for granting access to his private Mathematica–Singular interface and to his implementation of the algorithm for identifying non-vanishing sectors.
We wish to acknowledge Jonathan Ronca and William J. Torres Bobadilla for collaboration on related work, and Nicola Bartolo and Angelo Ricciardone for  interesting discussion.
The computations for this work were performed on the Slarti and Hypatia clusters at the Max Planck Institute for Gravitational Physics in Potsdam, with additional computing and storage resources provided by the CloudVeneto initiative at the University of Padova and INFN.
G.B.'s research is supported by the Italian MIUR under contract 20223ANFHR (PRIN2022) and by the ERC (NOTIMEFORCOSMO, 101126304).
R.P.’s research is funded by the Deutsche Forschungsgemeinschaft (DFG, German Research Foundation), Projektnummer 417533893/GRK2575 ``Rethinking Quantum Field Theory''.
M.P.'s research is supported by the European Union under the Next Generation EU programme.
G.B., M.M., P.M., M.P. and S.S. research has been supported by the INFN initiatives \textit{Amplitudes}, \textit{InDark}, and \textit{TPPC}.
R.P. and J.S. are partly supported by ERC grant (GWSky/ 101167314).

%%%%%%%%%%%%%% Appendices

\appendix
\setcounter{secnumdepth}{1}

\section{Potential in terms of connected correlators}
\label{app:potential_correlators}
{

We report the explicit expression for the static potential in terms of the connected correlation functions $\Gamma_{n_1,n_2}$, obtained expanding Eq.~\eqref{eq_all_order_V}.
The prime contributions $\Gamma_{A,B}$ in the sector $m_1^A m_2^B$
are highlighted in {\color{red}{red}}. These are the only genuinely new contributions that must be evaluated at a given PN order, and, by virtue of the factorization theorem, they appear exclusively at even PN orders.
Up to 7PN order the potential reads:
\begin{widetext}
\begin{equation}
\mathcal{V}_{\rm 0PN}^{G_N} = -m_1 m_2\textcolor{red} {\Gamma_{1,1}} \, , 
\end{equation}
\begin{equation}
\mathcal{V}_{\rm 1PN}^{G_N^2} = \frac{1}{2} m_1 m_2^2 \Gamma_{1,1}^2
+(m_1 \leftrightarrow m_2) \, , 
\end{equation}
\begin{equation}
\mathcal{V}_{\rm 2PN}^{G_N^3} = 
-m_1 m_2^3 \left(\frac{1}{6} \Gamma_{1,1}^3+\frac{1}{6} \textcolor{red}{\Gamma_{1,3}}\right)
-\frac{m_2^2 m_1^2}{2} \left(\Gamma_{1,1}^3+\frac{1}{4} \textcolor{red}{\Gamma_{2,2}}\right)
+ (m_1 \leftrightarrow m_2) \, ,
\end{equation}
\begin{equation}
\begin{split}
    \mathcal{V}_{\rm 3PN}^{G_N^4} &= m_1 m_2^4 \left(\frac{1}{24} \Gamma_{1,1}^4+\frac{1}{6} \Gamma_{1,3} \Gamma_{1,1}\right) 
     +m_1^2 m_2^3 \left(\Gamma_{1,1}^4+\frac{1}{2} \Gamma_{1,3} \Gamma_{1,1}+\frac{1}{2} \Gamma_{2,2} \Gamma_{1,1}\right)
+ (m_1 \leftrightarrow m_2) \, , 
\end{split}
\end{equation}
\begin{equation}
\begin{split}
    \mathcal{V}_{\rm 4PN}^{G_N^5} &=  m_1 m_2^5 \left(-\frac{1}{120} \Gamma_{1,1}^5-\frac{1}{12} \Gamma_{1,3} \Gamma_{1,1}{}^2-\frac{1}{120} \textcolor{red}{\Gamma_{1,5}}\right) \\ 
& +m_1^2 m_2^4 \left(-\frac{2}{3} \Gamma_{1,1}{}^5-\frac{7}{6} \Gamma_{1,3} \Gamma_{1,1}{}^2-\frac{1}{2} \Gamma_{2,2} \Gamma_{1,1}{}^2-\frac{1}{48} \textcolor{red}{\Gamma_{2,4}}\right) \\ 
& +\frac{m_1^3 m_2^3}{2} \left(-\frac{9}{4} \Gamma_{1,1}{}^5-\frac{3}{2} \Gamma_{1,3} \Gamma_{1,1}{}^2-2 \Gamma_{2,2} \Gamma_{1,1}{}^2-\frac{1}{36} \textcolor{red}{\Gamma_{3,3}}\right) \\ 
& +(m_1 \leftrightarrow m_2)\, ,
\end{split}
\end{equation}
\begin{equation}
\begin{split}
    \mathcal{V}_{\rm 5PN}^{G_N^6} &=  m_1 m_2^6 \left(\frac{1}{720} \Gamma_{1,1}^6+\frac{1}{36} \Gamma_{1,3} \Gamma_{1,1}^3+\frac{1}{120} \Gamma_{1,5} \Gamma_{1,1}+\frac{1}{72} \Gamma_{1,3}^2\right) \\ 
& +m_1^2 m_2^5 \left(\frac{1}{3} \Gamma_{1,1}^6+\frac{4}{3} \Gamma_{1,3} \Gamma_{1,1}^3+\frac{1}{3} \Gamma_{2,2} \Gamma_{1,1}^3+\frac{1}{24} \Gamma_{1,5} \Gamma_{1,1}+\frac{1}{24} \Gamma_{2,4} \Gamma_{1,1}+\frac{1}{8} \Gamma_{1,3}^2+\frac{1}{12} \Gamma_{1,3} \Gamma_{2,2}\right) \\ 
& +m_1^3 m_2^4 \left(3 \Gamma_{1,1}^6+\frac{49}{12} \Gamma_{1,3} \Gamma_{1,1}^3+\frac{15}{4} \Gamma_{2,2} \Gamma_{1,1}^3+\frac{1}{12} \Gamma_{2,4} \Gamma_{1,1}+\frac{1}{12} \Gamma_{3,3} \Gamma_{1,1}+\frac{1}{8} \Gamma_{2,2}^2+\frac{1}{4} \Gamma_{1,3} \Gamma_{2,2}+\frac{1}{12} \Gamma_{1,3}^2\right) 
\\ 
& 
+(m_1 \leftrightarrow m_2) \, ,
\end{split}
\end{equation}
\begin{equation}
\begin{split}
    \mathcal{V}_{\rm 6PN}^{G_N^7} &= m_1 m_2^7 \left(-\frac{\Gamma_{1,1}^7}{5040}-\frac{1}{144} \Gamma_{1,3} \Gamma_{1,1}^4-\frac{1}{240} \Gamma_{1,5} \Gamma_{1,1}^2-\frac{1}{72} \Gamma_{1,3}^2 \Gamma_{1,1}-\frac{\textcolor{red}{\Gamma_{1,7}}}{5040}\right) \\ 
& +m_1^2 m_2^6 \left(-\frac{2}{15} \Gamma_{1,1}^7-\Gamma_{1,3} \Gamma_{1,1}^4-\frac{1}{6} \Gamma_{2,2} \Gamma_{1,1}^4-\frac{11}{120} \Gamma_{1,5} \Gamma_{1,1}^2-\frac{1}{24} \Gamma_{2,4} \Gamma_{1,1}^2-\frac{5}{12} \Gamma_{1,3}^2 \Gamma_{1,1}-\frac{1}{6} \Gamma_{1,3} \Gamma_{2,2} \Gamma_{1,1}-\frac{\textcolor{red}{\Gamma_{2,6}}}{1440}\right) \\ 
& +m_1^3 m_2^5 \biggl(-\frac{45}{16} \Gamma_{1,1}^7-\frac{119}{16} \Gamma_{1,3} \Gamma_{1,1}^4-\frac{9}{2} \Gamma_{2,2} \Gamma_{1,1}^4 -\frac{5}{48} \Gamma_{1,5} \Gamma_{1,1}^2-\frac{7}{24} \Gamma_{2,4} \Gamma_{1,1}^2-\frac{1}{8} \Gamma_{3,3} \Gamma_{1,1}^2-\frac{7}{8} \Gamma_{1,3}^2 \Gamma_{1,1} 
\\ 
&\qquad \qquad  
-\frac{3}{8} \Gamma_{2,2}^2 \Gamma_{1,1}-\frac{3}{2} \Gamma_{1,3} \Gamma_{2,2} \Gamma_{1,1}-\frac{1}{720} \textcolor{red}{\Gamma_{3,5}}\biggr) \\ 
& +\frac{m_1^4 m_2^4}{2} \biggl(-\frac{64}{9} \Gamma_{1,1}^7-\frac{104}{9} \Gamma_{1,3} \Gamma_{1,1}^4-12 \Gamma_{2,2} \Gamma_{1,1}^4-\frac{1}{3} \Gamma_{2,4} \Gamma_{1,1}^2-\frac{5}{12} \Gamma_{3,3} \Gamma_{1,1}^2 -\frac{5}{4} \Gamma_{2,2}^2 \Gamma_{1,1}
\\ 
& \qquad \qquad -2 \Gamma_{1,3} \Gamma_{2,2} \Gamma_{1,1}-\frac{17}{18} \Gamma_{1,3}^2 \Gamma_{1,1}-\frac{1}{576} \textcolor{red}{\Gamma_{4,4}}\biggr) 
\\
& + (m_1 \leftrightarrow m_2)  \, , 
\end{split}
\end{equation}
\allowdisplaybreaks[4]
\begin{align}
    \mathcal{V}_{\rm 7PN}^{G_N^8} &= m_1 m_2^8 \left(\frac{\Gamma_{1,1}^8}{40320}+\frac{1}{720} \Gamma_{1,3} \Gamma_{1,1}^5+\frac{1}{720} \Gamma_{1,5} \Gamma_{1,1}^3+\frac{1}{144} \Gamma_{1,3}^2 \Gamma_{1,1}^2+\frac{\Gamma_{1,7} \Gamma_{1,1}}{5040}+\frac{1}{720} \Gamma_{1,3} \Gamma_{1,5}\right)  
    \notag
\\ 
& +m_1^2 m_2^7 \biggl(\frac{2}{45} \Gamma_{1,1}^8+\frac{5}{9} \Gamma_{1,3} \Gamma_{1,1}^5+\frac{1}{15} \Gamma_{2,2} \Gamma_{1,1}^5+\frac{1}{10} \Gamma_{1,5} \Gamma_{1,1}^3+\frac{1}{36} \Gamma_{2,4} \Gamma_{1,1}^3+\frac{11}{18} \Gamma_{1,3}^2 \Gamma_{1,1}^2+\frac{1}{6} \Gamma_{1,3} \Gamma_{2,2} \Gamma_{1,1}^2
\notag
\\ 
& \qquad \qquad 
+\frac{1}{720} \Gamma_{1,7} \Gamma_{1,1}+\frac{1}{720} \Gamma_{2,6} \Gamma_{1,1}+\frac{1}{48} \Gamma_{1,3} \Gamma_{1,5}+\frac{1}{240} \Gamma_{1,5} \Gamma_{2,2}+\frac{1}{144} \Gamma_{1,3} \Gamma_{2,4}\biggr) 
\notag
\\ 
& 
+m_1^3 m_2^6 \biggl(\frac{81}{40} \Gamma_{1,1}^8+\frac{63}{16} \Gamma_{2,2} \Gamma_{1,1}^5+\frac{747}{80} \Gamma_{1,3} \Gamma_{1,1}^5+\frac{2}{5} \Gamma_{1,5} \Gamma_{1,1}^3+\frac{1}{2} \Gamma_{2,4} \Gamma_{1,1}^3+\frac{1}{8} \Gamma_{3,3} \Gamma_{1,1}^3
\notag
\\ 
& \qquad \qquad 
+\frac{9}{16} \Gamma_{2,2}^2 \Gamma_{1,1}^2+\frac{29}{8} \Gamma_{1,3} \Gamma_{2,2} \Gamma_{1,1}^2+\frac{27}{8} \Gamma_{1,3}^2 \Gamma_{1,1}^2+\frac{1}{240} \Gamma_{2,6} \Gamma_{1,1}+\frac{1}{240} \Gamma_{3,5} \Gamma_{1,1}
\notag
\\ 
& \qquad \qquad 
+\frac{1}{48} \Gamma_{1,5} \Gamma_{2,2}+\frac{1}{24} \Gamma_{1,3} \Gamma_{2,4}+\frac{1}{48} \Gamma_{2,2} \Gamma_{2,4}+\frac{1}{240} \Gamma_{1,5} \Gamma_{1,3}+\frac{1}{72} \Gamma_{1,3} \Gamma_{3,3}
\biggr) \notag \\ 
& 
+m_1^4 m_2^5 \biggl(\frac{100}{9} \Gamma_{1,1}^8+\frac{239}{9} \Gamma_{1,3} \Gamma_{1,1}^5+\frac{70}{3} \Gamma_{2,2} \Gamma_{1,1}^5+\frac{25}{144} \Gamma_{1,5} \Gamma_{1,1}^3+\frac{167}{144} \Gamma_{2,4} \Gamma_{1,1}^3
\notag \\ 
& \qquad \qquad 
+\Gamma_{3,3} \Gamma_{1,1}^3+\frac{721}{144} \Gamma_{1,3}^2 \Gamma_{1,1}^2+\frac{9}{2} \Gamma_{2,2}^2 \Gamma_{1,1}^2+\frac{233}{24} \Gamma_{1,3} \Gamma_{2,2} \Gamma_{1,1}^2
\notag \\ 
& \qquad \qquad 
+\frac{1}{144} \Gamma_{3,5} \Gamma_{1,1}+\frac{1}{144} \Gamma_{4,4} \Gamma_{1,1}+\frac{1}{24} \Gamma_{2,2} \Gamma_{2,4}+\frac{1}{144} \Gamma_{1,5} \Gamma_{1,3}+\frac{5}{144} \Gamma_{2,4} \Gamma_{1,3}
\notag \\ 
& \qquad \qquad 
+\frac{1}{24} \Gamma_{1,3} \Gamma_{3,3}+\frac{1}{24} \Gamma_{2,2} \Gamma_{3,3}
\biggr) \notag \\ 
& + (m_1 \leftrightarrow m_2) \, .
\end{align}
\end{widetext}

}

\section{Potential at 8PN and 9PN}
\label{app:8_9_PN}

The master formula in Eq.~\eqref{eq_all_order_V}
can be used to derive the static contributions to higher PN orders in terms of correlation functions. In particular, after substituting the all-order value for the 0SF correlator, $\Gamma_{1,n}=\Gamma_{n,1}=(G_N/r)^n(n-1)!$ for $n$ odd, the $8$PN and $9$PN static potentials are fixed up to four undetermined correlation functions, which can be evaluated from explicit diagrammatic calculation at $8$-loops.
Defining the rescaled correlators
$
%\begin{equation}
%    \label{eq:rescaled_gamma_bar}
    \bar{\Gamma}_{n_1 , n_2} = 
%    \left(\frac{G_N}{r}\right)
(G_N/r)^{1 - n_1 - n_2} \, 
\Gamma_{n_1,n_2}\, ,
%\end{equation}
$
the static potential at $8$PN and $9$PN order, respectively, read as: 
\begin{widetext}
\begin{align}
 \mathcal{V}_{8{\rm PN}}^{G_N^9} & =  - \frac{G_N^9}{r^9}   \left( \frac{35}{128} m_1 m_2^9 +  \left(\frac{\textcolor{red}{\bar{\Gamma}_{2,8}}}{80640}+\frac{266}{9}\right) m_1^2 m_2^8 +\left(\frac{\textcolor{red}{\bar{\Gamma}_{3,7}}}{30240}+\frac{250705}{288}\right)  m_1^3 m_2^7  \right. \nonumber \\
 & \qquad \qquad\left. + \left(\frac{\textcolor{red}{\bar{\Gamma}_{4,6}}}{17280}+\frac{20084}{3}\right) m_1^4 m_2^6 + \left(\frac{\textcolor{red}{\bar{\Gamma}_{5,5}}}{14400}+\frac{7656577}{576}\right) \frac{m_1^5 m_2^5}{2}  \right) +(m_1\leftrightarrow m_2) \ ,
   \\
  \mathcal{V}_{9{\rm PN}}^{G_N^{10}} & =  \frac{G_N^{10}}{r^{10}} \left( \frac{63}{256} m_1 m_2^{10} + \left(\frac{\bar{\Gamma}_{2,8}}{40320}+\frac{577}{18}\right) m_1^2 m_2^9 +  \left(\frac{\bar{\Gamma}_{2,8}}{10080}+\frac{\bar{\Gamma}_{3,7}}{10080}+\frac{186233}{144}\right) m_1^3 m_2^8 \right. \nonumber \\ 
  & \left. \qquad \qquad  +  \left(\frac{\bar{\Gamma}_{3,7}}{4320}+\frac{\bar{\Gamma}_{4,6}}{4320}+\frac{270781}{18}\right) m_1^4 m_2^7 + \left(\frac{\bar{\Gamma}_{4,6}}{2880}+\frac{\bar{\Gamma}_{5,5}}{2880}+\frac{2501525}{48}\right) m_1^5 m_2^6 \right)+(m_1\leftrightarrow m_2) \, ,
\end{align}
\end{widetext}
where  
the presently \textit{unknown} 8PN correlation functions are indicated in {\color{red}{red}}. 
At 9PN, no new correlation functions appear.
This is consistent with the fact that new prime correlation functions enter only at even-PN orders, whereas the odd-PN static potentials are fixed recursively in terms of lower-order correlation functions.

\section{All-order values of $0$SF correlators}
\label{app:test_body}
We present the derivation of the all-order expression for the
$0$SF correlators $\Gamma_{1,n}=\Gamma_{n,1}$, by matching the general
correlation-function formula of Eq.~\eqref{eq_all_order_V}
to the exact test-body limit.

We focus on the sector of the static potential linear in $m_1$,
which corresponds to the $0$SF contribution. 
Expanding
Eq.~\eqref{eq_all_order_V} to first order in $m_1$, one obtains
\begin{align}
\cV_{\rm static}^{0{\rm SF}}
&=
m_1\lim_{\hbar\to0}
\Bigg[
e^{\frac{\ii m_2}{\hbar}}
\sum_{b\ge0}\frac{1}{b!}
\left(-\frac{\ii m_2}{\hbar}\right)^b
\notag\\
&\times
\exp\!\left(
\sum_{\substack{n\ge1\\ n\,  \mathrm{odd}}}
\frac{(-\ii \hbar\, b)^{n}}{n!}\,\Gamma_{1,n}
\right)
\Bigg] ,
\label{eq:app_0SF_start}
\end{align}
where we restricted the sum to $\Gamma_{1,n}$ since they are the only correlation functions contributing to the 0SF potential.
To evaluate the classical limit, we define
\begin{equation}
    f(x) = \exp\!\left(\sum_{\substack{n\ge1\\ n\,  \mathrm{odd}}} \frac{x^{n}}{n!}\,\Gamma_{1,n} \right)\ .
\end{equation}
Then, recalling the identity $\exp\!\left(\alpha \frac{\partial}{\partial x}\right) f(x) = f(x + \alpha)$, with $\alpha = - \ii \hbar$ it holds

\begin{align}
   &e^{\frac{m_2}{\alpha}}
\sum_{b\ge0}\frac{1}{b!}
\left(-\frac{m_2}{\alpha}\right)^b  f( \alpha b ) \nonumber \\
% &= e^{\frac{m_2}{\alpha}}
% \sum_{b\ge0}\frac{1}{b!}
% \left(-\frac{m_2}{\alpha}\right)^b  f( x + \alpha b ) \biggr|_{x = 0}\\
&= e^{\frac{m_2}{\alpha}}
\sum_{b\ge0}\frac{1}{b!}
\left(-\frac{m_2}{\alpha}\right)^b \exp\left(\alpha b \frac{\partial}{\partial x}\right) f(x) \biggr|_{x = 0} \\
%
% &= e^{\frac{m_2}{\alpha}}
% \sum_{b\ge0}\frac{1}{b!}
% \left(-\frac{m_2}{\alpha} \exp\left(\alpha \frac{\partial}{\partial x}\right)\right)^b f(x) \biggr|_{x = 0} \\
&= \exp\left( -\frac{m_2}{\alpha} \left( \exp\left(\alpha \frac{\partial}{\partial x}\right) - 1\right)\right) f(x) \biggr|_{x = 0} 
\nonumber \\
% &= \exp\left( - m_2  \frac{\partial}{\partial x} \right) f(x) + \mathcal{O}(\alpha) \biggr|_{x = 0} \\
&= f(- m_2) + \mathcal{O}(\alpha) \ , \nonumber
\end{align}
where in the last equality we expanded the exponential in $\alpha$, obtaining $\exp\left( - m_2  \frac{\partial}{\partial x} \right)+ \mathcal{O}(\alpha) $.

Since $\alpha \propto \hbar$, and $\Gamma_{1,n} \propto \hbar^0$, the classical limit in Eq.~\eqref{eq:app_0SF_start} reduces to:
\begin{equation}
\cV_{\rm static}^{0{\rm SF}}
=
m_1
\exp\!\left[
-\sum_{\substack{n\ge1\\ n\ {\rm odd}}}
\frac{m_2^n}{n!}\,\Gamma_{1,n}
\right]\,  .
\label{eq:app_0SF_final}
\end{equation}
On the other hand, in the exact test-body limit it holds
\begin{align}
\cV_{m_1\ll m_2}
&=
m_1\sqrt{g_{00}}
=
m_1\sqrt{\frac{1-\frac{G_N m_2}{r}}{1+\frac{G_N m_2}{r}}}
\notag\\
&=
m_1
\exp\!\left[
-\sum_{\substack{n\ge1\\ n\ {\rm odd}}}
\frac{1}{n}\left(\frac{G_N m_2}{r}\right)^n
\right]\,  ,
\label{eq:app_test_body}
\end{align}
where in the last step we used
$\sum_{n\ge1,\;n\ {\rm odd}} z^n/n
=\frac12\log\!\left(\frac{1+z}{1-z}\right)$.
Matching Eq.~\eqref{eq:app_0SF_final}
and Eq.~\eqref{eq:app_test_body} yields
\begin{align}
\Gamma_{1,n}
&=
\left(\frac{G_N}{r}\right)^n (n-1)!\  \quad \text{for }n\text{ odd}\ .
\end{align}

\bibliography{biblio}

\end{document}